\documentclass[amsmath, amssymb, twocolumn, floatfix, superscriptaddress, nofootinbib]{revtex4-2}

\usepackage{braket}
\usepackage{graphicx}
\usepackage{enumitem}
\usepackage{siunitx}
\usepackage{booktabs}

\usepackage{placeins}
\usepackage{float}
\usepackage{tikz}

\usepackage{MnSymbol} 
\newcommand{\blackdiam}{\rotatebox[origin=c]{45}{$\blacksquare$}}

\usepackage{mathtools}
\usepackage{ulem} 

\makeatletter
\newcommand{\globalcolor}[1]{%
  \color{#1}\global\let\default@color\current@color
}

\makeatother

\makeatletter
\def\p@subsection{}

\makeatother

\newcommand*\circmarker{\tikz{\node[shape=circle,draw,inner sep=0pt,minimum size=2mm] {}}}
\newcommand*\squaremarker{\tikz{\node[shape=rectangle,draw,inner sep=0pt,minimum size=1.9mm] {}}}

\newcommand{\PRLsep}{\noindent\makebox[\linewidth]{\resizebox{0.625\linewidth}{1pt}{$\bullet$}}\bigskip}

\begin{document}

\hyphenpenalty9999
\hbadness=99999

\title{\Large{Experimental Observation of Earth's Rotation with Quantum Entanglement}}

\author{Raffaele~Silvestri}\affiliation{University of Vienna, Faculty of Physics \& Vienna Doctoral School in Physics,  Boltzmanngasse 5, A-1090 Vienna, Austria}
\author{Haocun~Yu}\email[Corresponding author: ]{haocun.yu@univie.ac.at}\affiliation{University of Vienna, Faculty of Physics \& Research Network Quantum Aspects of Space Time (TURIS), Boltzmanngasse 5, 1090 Vienna, Austria}
\author{Teodor~Strömberg}\affiliation{University of Vienna, Faculty of Physics \& Vienna Doctoral School in Physics,  Boltzmanngasse 5, A-1090 Vienna, Austria}
\author{Christopher~Hilweg}\affiliation{University of Vienna, Faculty of Physics \& Research Network Quantum Aspects of Space Time (TURIS), Boltzmanngasse 5, 1090 Vienna, Austria}
\author{Robert~W.~Peterson}\affiliation{University of Vienna, Faculty of Physics \& Research Network Quantum Aspects of Space Time (TURIS), Boltzmanngasse 5, 1090 Vienna, Austria}
\author{Philip~Walther}\email[Corresponding author: ]{philip.walther@univie.ac.at}\affiliation{University of Vienna, Faculty of Physics \& Research Network Quantum Aspects of Space Time (TURIS), Boltzmanngasse 5, 1090 Vienna, Austria}

\date{\today}

\begin{abstract}
Precision interferometry with quantum states has emerged as an essential tool for experimentally answering fundamental questions in physics. Optical quantum interferometers are of particular interest due to mature methods for generating and manipulating quantum states of light. The increased sensitivity offered by these states promises to enable quantum phenomena, such as entanglement, to be tested in unprecedented regimes where tiny effects due to gravity come into play. However, this requires long and decoherence-free processing of quantum entanglement, which has not yet been explored for large interferometric areas. Here we present a table-top experiment using maximally path-entangled quantum states of light in an interferometer with an area of \SI{715}{\meter^2}, sensitive enough to measure the rotation rate of Earth. A rotatable setup and an active area switching technique allow us to control the coupling of Earth’s rotation to an entangled pair of single photons. The achieved sensitivity of \SI{5}{\micro~\radian~\s^{-1}} constitutes the highest rotation resolution ever achieved with optical quantum interferometers, surpassing previous work by three orders of magnitude~\cite{Fink_2019}. Our result demonstrates the feasibility of extending the utilization of maximally entangled quantum states to large-scale interferometers. Further improvements to our methodology will enable measurements of general-relativistic effects on entangled photons~\cite{Brady_2021} opening the way to further enhance the precision of fundamental measurements to explore the interplay between quantum mechanics and general relativity along with searches for new physics~\cite{Stedman_1997}.
\end{abstract}

\maketitle

\section{Introduction}
For more than a century, interferometers have been important instruments to experimentally test fundamental physical questions. They disproved the luminiferous ether, helped to establish special relativity~\cite{Michelson_1887, MGP_1925} and enabled the measurement of tiny ripples in space-time itself known as gravitational waves~\cite{LIGO_2016}. With recent advances in technology interferometers can nowadays also operate using various different quantum systems like electrons~\cite{Zimmerman_1965, Hasselbach_1993}, neutrons~\cite{Werner_1979},
atoms~\cite{Riehle_1991, Lenef_1997, Gustavson_2000, Stockton_2011, Gautier_2022}, superfluids~\cite{Schwab_1997,Simmonds_2001}, and Bose-Einstein condensates~\cite{Gupta_2005,Levy_2007,Marti_2015}.
Quantum interferometers are of interest for two main reasons: First, the exploitation of quantum entanglement allows for super-resolving phase measurements that go beyond the standard quantum limit~\cite{Lee_2002, Giovannetti_2006}.
Second, the enhanced sensitivity of quantum interferometers opens up opportunities for precision measurements that can explore new frontiers in physics. These include setting constraints on dark energy models~\cite{Jaffe_2017}, testing quantum phenomena in non-inertial reference frames~\cite{Restuccia_2019, Cromb_2023, Fink_2019}, and investigating the interplay between quantum mechanics and general relativity~\cite{Amelino_1999, Bosi_2011, Yair_2021, Asenbaum_2017, Asenbaum_2020}.

Optical systems are particularly well-suited for realizing quantum interferometers, thanks to mature techniques available for generating a variety of quantum states, ranging from squeezed vacuum~\cite{ligo_squeezing, Yu_2020, virgo_squeezing, Mehmet_2010} to maximally path-entangled photons~\cite{Mitchell_2004, Nagata_2007}.
The latter are referred to as N00N states, represented by $\frac{1}{\sqrt{N}}(\Ket {N}_{a} \Ket{0}_{b} - \Ket {0}_{a} \Ket{N}_{b})$, 
wherein $N$ photons exist in a superposition of $N$ photons in mode $a$ with zero particles in mode $b$, and vice versa~\cite{Lee_2002}. Remarkably, these states behave like those of a single photon with $N$ times the energy, enabling a phase sensitivity at the Heisenberg limit that scales as $1/N$, and thus goes beyond the $1/\sqrt{N}$ scaling of the
standard quantum limit~\cite{Giovannetti_2006}.
Another advantage of photonic systems is that fiber-optical interferometers offer a clear pathway for expanding the interferometric area while maintaining a low level of quantum decoherence.

\begin{figure*}[!thb]
\centering

\includegraphics[width=1.0\textwidth]{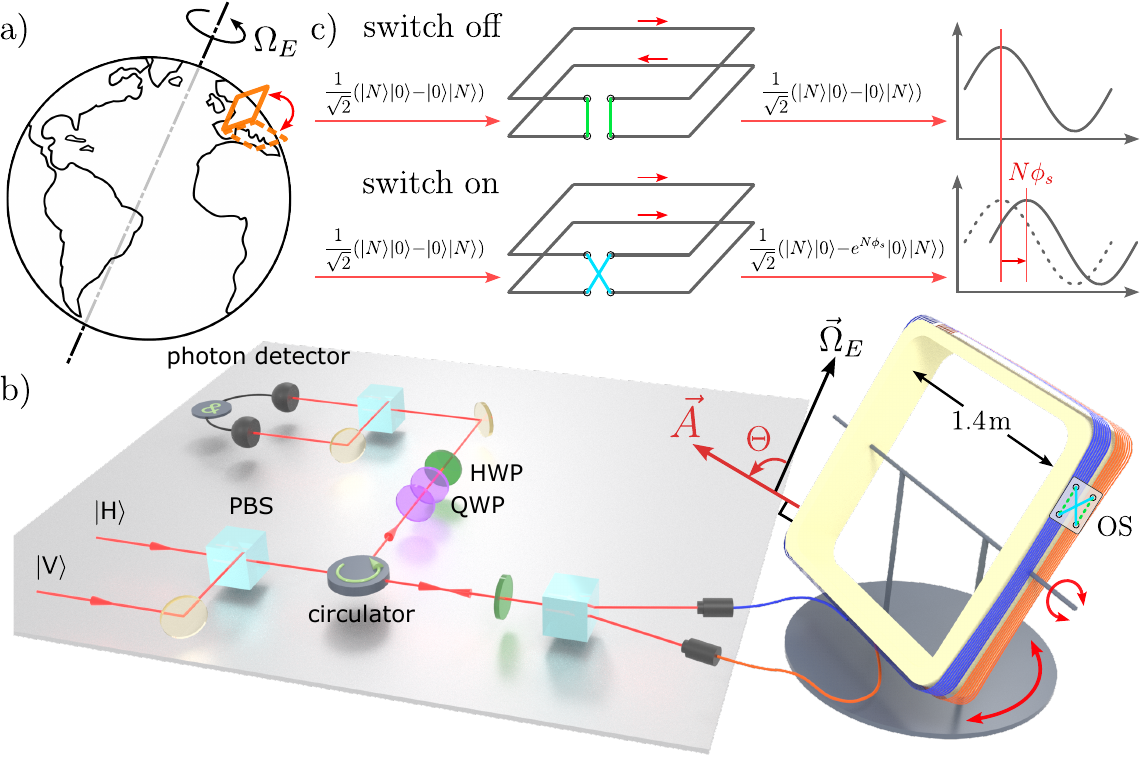}
\caption{\textbf{Earth's rotation measured using entangled photons.} 
\textbf{a}, A rotatable \SI{715}{\meter\squared} Sagnac fiber interferometer is built in a laboratory located in Vienna, Austria. 
\textbf{b}, Simplified schematic of the experimental setup. 
Orthogonally polarized photon pairs are converted to path-entangled N00N states in the Sagnac interferometer via a half-wave plate (HWP) followed by a polarizing beam splitter (PBS).
The frame angle $\Theta$ is defined as the angle between Earth's angular velocity vector $\vec{\Omega}_{E}$ and the fiber loop area vector $\vec{A}$. The signal is extracted by observing the phase shift of quantum interference fringes induced by Earth's rotation, using a set of quarter-wave plates (QWP) and a HWP, in combination with single photon coincidence counting (\&).
\textbf{c}, An optical switch (OS) is used to toggle Earth's rotation signal on and off independent of the frame angle $\Theta$. This is achieved by controlling the propagation direction (clockwise or counterclockwise) of photons in one half of the fiber spool.}
\label{fig:layout}
\end{figure*}

In this work, we report an experiment measuring the rotation of the Earth harnessing quantum entanglement in a large-scale optical fiber interferometer. 
We inject two-photon N00N states into a \SI{715}{\meter\squared} Sagnac interferometer, using quantum interference to demonstrate super-resolution while extracting Earth’s rotation rate. This goes beyond previous laboratory demonstrations of measurements probing Sagnac interferometers with quantum states of light that involved 
centimeter-scale fiber interferometers with at most hundred-meter-length fibers~\cite{Restuccia_2019, Cromb_2023, Fink_2019, Mehmet_2010, Bertocchi_2006}, and which were only used to measure synthetic signals.
We are able to confirm an acquired Sagnac phase from Earth's rotation with an enhancement factor of two due to the two-photon entangled state, achieving a rotation sensitivity of a few \SI{}{\micro\radian\per\second}. 
To the best of our knowledge, this is the largest quantum-optical Sagnac interferometer in the world, surpassing previous state-of-the-art rotation sensors employing two-particle entanglement. This measurement represents a significant milestone in the development of larger-scale quantum interferometers.

\section{Quantum optical Sagnac interferometer}
\begin{figure*}[!thb]
\centering
\includegraphics[width=1.0\textwidth]{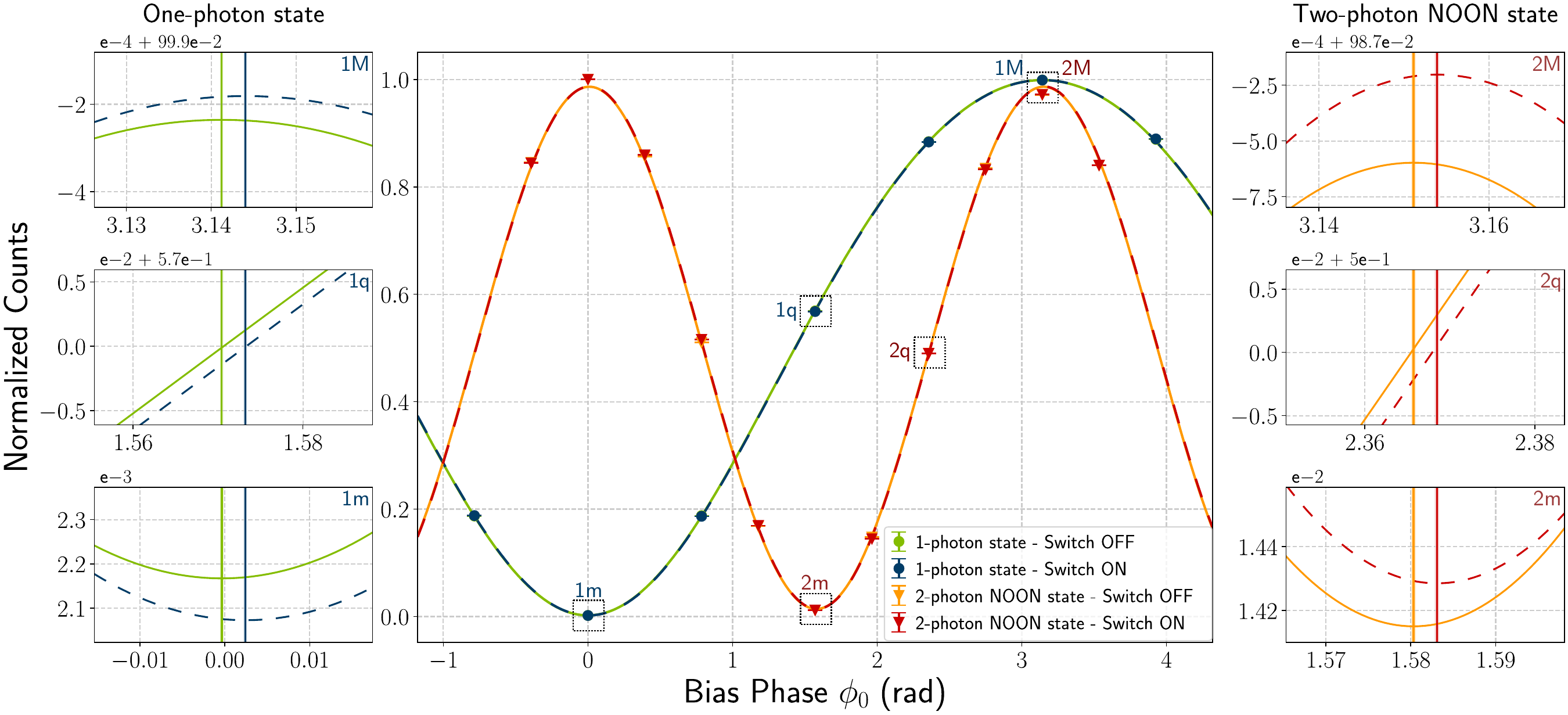}
\caption{
\textbf{Quantum interference measurement revealing the Sagnac phase shift induced by Earth's rotation.} 
\textbf{Center}, normalized quantum interference fringes of single-photon and two-photon entangled state measurements. 
The red and orange (blue and green) marks show the normalised two- (one-)photon coincidence counts with the Earth rotation signal switched on and off, respectively. The corresponding curves are least-squares fits to the data using a model of the experiment (see Appendix B). The doubled fringe frequency of the two-photon curves reveals the super-resolution due to quantum entanglement. \textbf{Sides}, Sagnac phase shifts induced by Earth's rotation at $\Theta = 2.5^\circ$, zooming in around $\phi = \pi, \pi/2, 0$ for single photon measurement (left), and around $\phi = \pi, 3\pi/4, \pi/2$ for two-photon measurement (right). The widths of the vertical lines indicate the size of uncertainties due to uncorrelated photon counting noise.
Because the same phase bias $\phi_0$ has been applied to both one-photon and two-photon measurements, the doubled Sagnac phase shift does not manifest in the plots. 
1: one-photon sate; 2: two-photon N00N state; M: maximum; m: minimum; q: quadrature.
}\label{fig:interference}
\end{figure*}

A considerable barrier to detecting Earth's rotation in large-scale fiber interferometers is its minute rate, fixed direction, and the inability to manipulate its behavior.
On the other hand, the ubiquitous presence of acoustic- and seismic vibrations and thermal fluctuations transduce directly into phase noise in optical fiber~\cite{Hilweg_22} and drive the motion of the large apparatus. To solve these problems we build our rotatable fiber interferometer with an optical switch to turn Earth's rotation signal on and off, allowing us to fully characterize the angle-dependent Sagnac phase (Fig.~1).


According to the Sagnac effect~\cite{Sagnac_RMP}, the flying times of photons traveling in opposite directions around a rotating encircled path are different, inducing a measurable phase difference:
\begin{align}
    \phi_{s} = \frac{8\pi\Omega_{E} A \cos{\Theta}}{\lambda c} .
    \label{eq: sagnac}
\end{align}  
Here, $\Omega_{E}$ is the rotation angular frequency of the Earth; $A$ is the interferometer's effective area of \SI{715}{\meter\squared} (for the calibration of the apparatus see details in Appendix A); $\Theta$ is the angle between the area vector of the fiber loop and the angular velocity vector of the Earth; $\lambda$ is the photon wavelength of \SI{1546}{\nano\meter}. 

In a Sagnac interferometer with an induced phase shift $\phi_s$, a coherent or a single-photon state, represented as $(\Ket {1}_{a}\Ket{0}_{b} - \Ket {0}_{a}\Ket{1}_{b})/\sqrt{2}$, evolves to 
$(\Ket {1}_{a}\Ket{0}_{b} - e^{i\phi_{s}}\Ket {0}_{a}\Ket{ 1}_{b})/\sqrt{2}$. 
After interference, a projective measurement at the output with two modes $a$ and $b$ gives rise to the probabilities of detecting a single photon $P_{1, a} = (1 + \cos(\phi_{s}))/2$ and $P_{1, b} = (1 - \cos(\phi_{s}))/2$ in each of the two output modes. 
Multi-photon interference occurs when we inject the entangled state
$(\Ket {N}_{a} \ket{0}_{b} - \Ket {0}_{a} \Ket{N}_{b})/\sqrt{2}$
into the interferometer, where the $N$ photons are in a superposition of being in either of the two modes. 
After transmission through the interferometer, the state evolves to 
$(\Ket {N}_{a} \ket{0}_{b} - e^{iN\phi_{s}}\Ket {0}_{a} \Ket {N}_{b})/\sqrt{2}$. 
This results in the probability of finding photons in two modes oscillating at $N$ times the frequency~$P_{N, a} = (1 + \cos(N\phi_{s}))/2$ and~$P_{N, b} = (1 - \cos(N\phi_{s}))/2$, enhancing the observed phase shift by a factor of $N$. 

\begin{figure*}[!htb]
\centering
\includegraphics[width=1.0\textwidth]{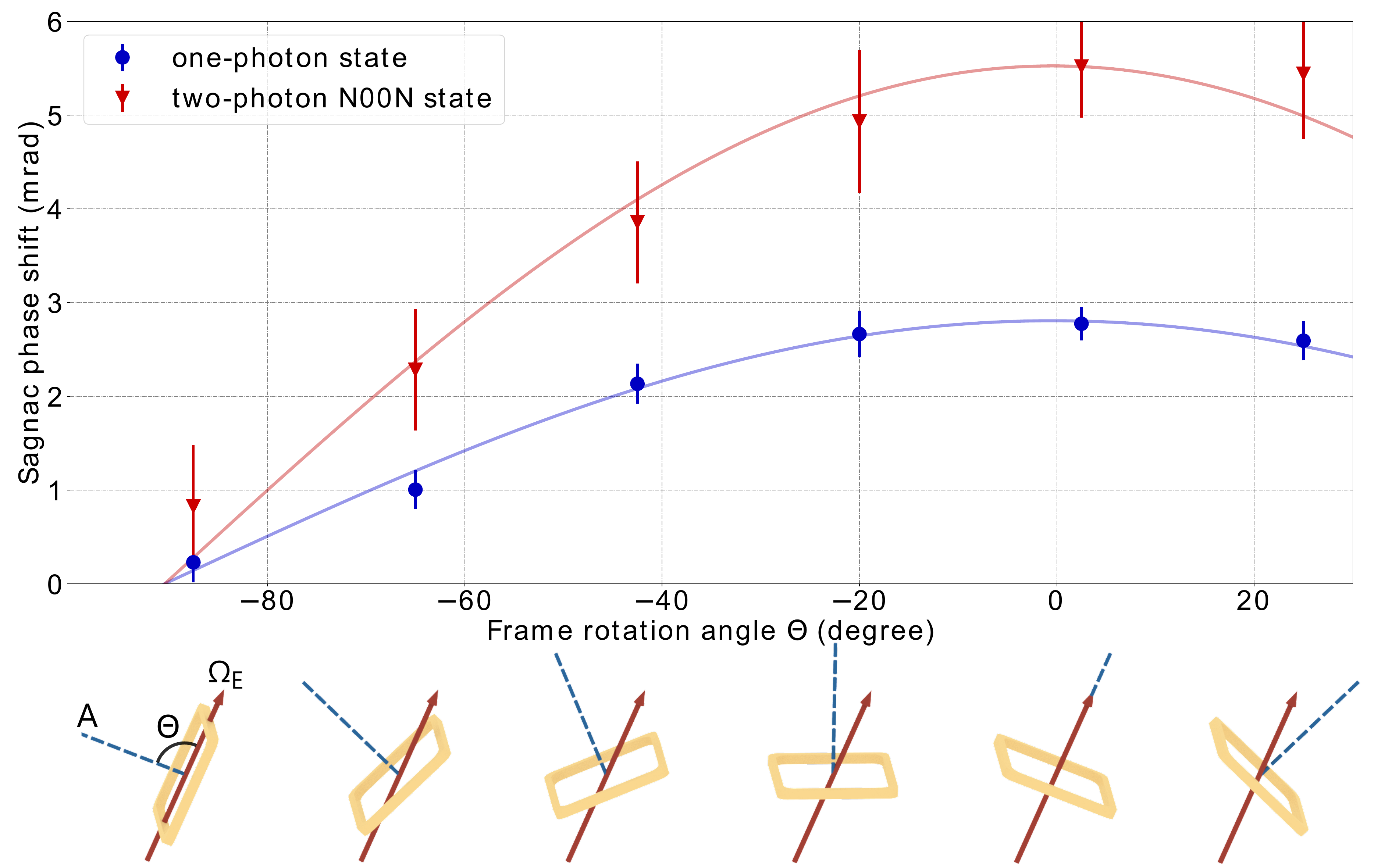}
\caption{\textbf{Sagnac phase shifts induced by Earth's rotation measured at six interferometer frame angles.} $\Theta$'s range from $-67.5^{\circ}$ to $+25^{\circ}$, evenly spaced by $22.5^{\circ}$.
\textbf{Top}, each data point is obtained with the same measurement sequence and extraction method as Fig.~\ref{fig:interference}. 
At each angle $\Theta$, the Sagnac phase shift measured using two-photon entangled N00N states (red triangle marks) is found to be doubled compared with single-photon states (blue circle marks).
The blue and red curves are the least-squares fits to  equation (\ref{eq: sagnac}) of the one-photon and two-photon N00N state measurements, respectively.
\textbf{Bottom}, representation of different angles between the area vector of the interferometer (blue line) and Earth's rotation angular velocity vector (red arrow).
The Sagnac phase shift induced by Earth's rotation can be increased and decreased as the frame rotation angle is varied.}
\label{fig: frame rot}
\end{figure*}


\section{Experimental implementation}
The two-photon path-entangled state is realized by exploiting the polarization correlation of photon pairs emitted by a type-II spontaneous parametric down-conversion (SPDC) source~\cite{Greganti_2018}. 
The photon pairs, centered at \SI{1546}{\nano\meter}, are created in the product state $\Ket{1}_H \Ket{1}_V$, where H and V denote horizontal and vertical polarization, respectively.
A half-wave plate (HWP) oriented at \SI{22.5}{\degree} (with respect to the horizontal axis) transforms this product state into the polarization-entangled two-photon N00N state $(\Ket{2}_H \Ket{0}_V - \Ket{0}_H \Ket{2}_V)/\sqrt{2}$, where the cross terms cancel out due to the indistinguishability of the photons.
Subsequently, this state is converted into a path-entangled state at a polarizing beamsplitter (PBS), which separates the H and V photons into clockwise and counter-clockwise propagating modes.
After passing through the \SI{2}{\kilo\meter} fiber loops, the clockwise-traveling photons pick up a Sagnac phase shift $\phi_{s}$ induced by Earth's rotation relative to the counterclockwise-traveling ones. The same PBS then converts the state back into the polarization-entangled state:
\begin{align}
    \frac{1}{\sqrt{2}}(\Ket {2}_H \Ket{0}_V - e^{i2\phi_{s}}\Ket {0}_H \Ket {2}_V) \,.
    \label{eq: state evolve}
\end{align}
Interference takes place again at the $22.5^\circ$ HWP, leading to the output state:
\begin{align*}
    \frac{1}{\sqrt{2}}\sin\phi_{s}(\Ket{2}_H \Ket{0}_V + \Ket{0}_H \Ket{2}_V) - i\cos\phi_{s}\Ket{1}_H \ket{1}_V.
\end{align*}
A set of waveplates is used to control the detection probabilities by introducing a bias phase $\phi_0$. 
This artificially adds a relative phase between the H and V polarization components, allowing us to scan the full interference fringe and also project the measurements onto any polarization basis, and turns the state into:
\begin{align} 
    \frac{1}{\sqrt{2}}(\Ket {2}_H \Ket{0}_V - e^{i(2\phi_{0}+2\phi_{s})}\Ket {0}_H \Ket{2}_V) .
    \label{eq: state evolve detection}
\end{align}
To perform a projective measurement onto the $\Ket{1}_H \Ket{1}_V$ component of the state, we analyze the two-fold coincidence probability $P_{HV}$ by collecting photons in both output ports of the PBS before detectors:
\begin{align}
    P_{HV} = \frac{1}{2} \left[1 + \cos(2\phi_0+2\phi_{s}) \right] .
    \label{eq: prob}
\end{align}
This gives an enhancement factor of two in the observed Sagnac phase, as well as in the bias phase.

The central component of the Sagnac interferometer consists of \SI{2}{\kilo\meter} fibers wound around a \SI{1.4}{\meter} square aluminum frame (yellow)~(Fig.~\ref{fig:layout} b). Because the detectable Sagnac phase shift induced by Earth's rotation depends on the direction of the area vector $\vec{A}$, the frame is designed to be rotatable in both pitch and yaw dimensions. This allows for a series of measurements to be taken at different values of $\Theta$. 

To more distinctly manifest the rotation signal, an optical switch is incorporated to toggle the effective area of the interferometer.
The optical fiber is divided into two equal \SI{1}{\kilo\meter} fiber segments (orange and blue), which are connected by the four-port optical switch. As shown in Fig.~\ref{fig:layout} c, flipping the optical switch reverses the direction of light propagation in one of the fiber loops. When the optical switch is in the “OFF” state, the Sagnac phase shift is canceled out due to the opposite directions of light propagation in the two fiber segments, resulting in two area vectors with opposite signs and a zero effective area. By comparing the measurements in the optical switch “ON” and “OFF” states, it can be confirmed that the observed phase shifts are exclusively caused by Earth's rotation.

From equation (\ref{eq: sagnac}), the Sagnac phase is maximized when the interferometer is oriented in a way that Earth's rotation vector perpendicularly intersects the plane of the interferometer area. This orientation is determined from a calibration procedure with classical light in the interferometer (see details in Appendix A).
Figure~\ref{fig:interference} shows the data for the Sagnac phase shifts induced by Earth’s rotation at $\Theta=2.5^\circ$. The data points are acquired for one- and two-photon N00N states propagating through the interferometer. For the two-photon entangled states, eleven different data points were taken while continuously switching between the two operating modes: with and without Earth’s rotation signal (switch on and off, respectively). When alternating operation between the two modes at a frequency of \SI{0.1}{\hertz}, Earth’s rotation signal is resolved by comparing the interference fringes of the two modes. To further confirm that the phase shift is solely due to Earth’s rotation, additional data are acquired at various frame angles, thereby enabling curve fitting and precise phase-difference extraction, as depicted in Fig.~\ref{fig: frame rot}.

\section{Results}
Fig.~\ref{fig:interference} shows quantum interference fringes of the Sagnac interferometer at $\Theta=2.5^\circ$. In the central figure, the red and orange marks represent normalized two-photon coincidence counts $P_{H}$ and $P_{V}$ measured with the optical switch on and off, respectively. 
These data were generated from eleven sets of 30-min contiguous integration periods. 
Each data set was taken with a specific value of $\phi_0$, ranging from $-\pi/8$ to $2\pi+\pi/8$ to cover a full interference fringe.
The blue and green marks are heralded single-photon measurements, with eleven (seven shown) 15-min contiguous integration periods, ranging from $-\pi/4$ to $2\pi+\pi/4$, serving as the reference measurement.
The uncertainties for each data point are represented by $\pm1$ standard deviations, which were calculated from Monte Carlo simulations using $10^{5}$ samples of Poisson-distributed photon coincidence counts (see details in Appendix A for a comprehensive error analysis).

For the two-photon measurements, the eleven data points obtained in each switch mode are fit to an interference fringe model. Earth's rotation signal is extracted by calculating the phase shift between the two curves (red and orange).
Based on equation~(\ref{eq: prob}), the data are fit with:
    \begin{align}
    N_{\text{switch off}}(\phi) & = N_{0}(1+\mathcal{V}\cos(2\phi)) \label{eq: fitting model off}\\
    N_{\text{switch on}}(\phi)  & = N_{0}(1+\mathcal{V}\cos(2\phi + \phi_{s}^{(2)})), 
      \label{eq: fitting model on}
    \end{align}
where ${N_{0}}$ is the amplitude of the photon interference, $\mathcal{V}$ is the interference visibility, and $\phi_{s}^{(2)}$ is Earth's rotation-induced phase shift to be measured.
The extracted phase difference between two interference fringes is $\phi_{s}^{(2)} = \SI{5.5\pm0.5}{\milli\radian}$.
A similar fitting and phase extraction procedure is employed for single-photon reference measurements (blue and green), resulting in $\phi_{s} = \SI{2.8\pm0.2}{\milli\radian}$. 
In the two-photon measurement, the phase shift is enhanced by a factor of two due to the presence of entanglement.

Sagnac phase shift measurements at five additional frame angles $\Theta$ are presented in Fig.~3. 
This plot explicitly shows two things: 
First, the Sagnac phase shift induced by Earth's rotation is proportional to $\cos(\Theta)$ as expected from equation~(\ref{eq: sagnac}). 
Second, the two-photon measurements consistently exhibit a doubled phase compared to the single-photon measurements for all the different frame angles.
For each value of $\Theta$, the Sagnac phase shift is extracted by comparing the interference fringes with the optical switch on and off, following a procedure identical to that used for $\Theta = 2.5^\circ$.
The data for the five additional angles were acquired with a shorter integration time compared to Fig.~\ref{fig:interference}, resulting in correspondingly larger statistical uncertainties.
The red and blue curves are the least-squares fits using equation (\ref{eq: sagnac}). 
From these fits, the maximum Sagnac phase shift induced by Earth's rotation in the two-photon N00N state is $\SI{5.5\pm0.4}{\milli\radian}$, which corresponds to an Earth's rotation rate of $\Omega_E = \SI{7.1\pm0.5e-5}{\radian/\s}$, compared with $\SI{2.8\pm0.1}{\milli\radian}$ or $\Omega_E = \SI{7.2\pm0.3e-5}{\radian/\s}$ in the one-photon measurement. Both agree with the internationally-accepted value \SI{7.3e-5}{\radian/\s}~\cite{Earthrotationrate}, leading to an enhancement factor of $\SI{1.97\pm0.16}{}$.


\section{Discussion and Conclusions:} 

We have demonstrated the largest and most precise quantum optical Sagnac interferometer to date, exhibiting sufficient sensitivity to measure Earth's rotation rate using both single and two-photon entangled states. When comparing the use of quantum entanglement to classical probes, we observed a factor of two improvement in the measured phase value due to super-resolution. Our approach is readily scalable to N00N states with higher photon numbers~\cite{Mitchell_2004}, with the main limitations being the large amount of transmission loss of the experimental setup and the generation rate of the multi-photon states. The achievable phase resolution is primarily hindered by the scale factor instability, with the most detrimental contributions coming from mechanical vibrations of the frame due to its extensive surface area, thermal fluctuations, and acoustic noise. 

However, a major advantage of our scheme, including the effective area-switching, is its compatibility with the current cutting-edge technology of fiber-optic gyroscopes (FOGs). FOGs have reached phase resolutions of less than a \SI{}{\nano\radian} translating to rotation rates below \SI{0.1}{\nano\radian~\s^{-1}}, with a stable signal over more than a month~\cite{CFOG_2014, LFOG_2020}. 

Remarkably, an analysis suggests that pairing these gyroscopes with state-of-the-art sources~\cite{Neumann_2022} of two-photon entangled states would lead to a four-orders-of-magnitude enhancement in sensitivity over our present work. Anticipating the improvements of future single-photon sources, a logical next step will be the use of the recently proposed giant-FOG with an area of \SI{15.000}{\meter^{2}}~\cite{GFOG_2017} reaching a resolution of about \SI{20}{\pico\radian~\s^{-1}}. However, for the observation of general-relativistic effects that couple to two-photon entanglement , such as frame-dragging and geodetic corrections~\cite{Brady_2021} to the Sagnac phase, a resolution below \SI{0.1}{\pico\radian~\s^{-1}} is needed.
We anticipate that a square fiber ring interferometer (GFRING) with an area on the order of \SI{20}{\kilo\meter^{2}} would be able to probe a regime lower than the general relativistic rotation rate correction term $\Omega_{GR}=10^{-9}\Omega_{E}$ (see Fig. 5).
Although interferometers such as ring laser gyroscopes hold the record as the most precise local rotation sensors, already being at the $\Omega_{GR}$ threshold~\cite{DiVirgilio_2022}, the feasibility of probing these ring cavities with entangled photon pairs is limited by the requirement of generating single-photons with ultranarrow linewidth at high efficiency.
We therefore believe that fiber ring interferometers are the most promising platform to investigate the same regimes with entangled photon states.
Our work paves the way for other technically challenging proposals, such as directly probing gravitationally-induced phase shifts~\cite{Zych_2012}, rotational and gravitational decoherence in the quantum interference of photons~\cite{Kish_2022}, testing fundamental symmetries in quantum field theory~\cite{Stedman_1997}, investigating local Lorentz invariance violation~\cite{Moseley_2019}, detecting exotic low-mass fields from high-energy astrophysical events~\cite{Dailey_2021}, and dark matter searches like axion-photon coupling~\cite{Stedman_1997}. Indeed, it is predicted that two photons can transition to axions in presence of an external magnetic field. In an optical Sagnac interferometer where two orthogonal polarizations counter-propagate, the component parallel to the magnetic field would then be retarded with respect to the other leading to a non-reciprocal observable phase shift or the loss of two photons. Our interferometer constitutes an excellent testbed for this phenomenon, allowing investigations both with classical light and entangled photons.
In conclusion, the successful observation of the effect of Earth's rotation highlights the practical feasibility of large-scale optical fiber interferometers with entangled photons. After a century from the first local observation of Earth's rotation-induced fringe shift with light in a Sagnac interferometer~\cite{MGP_1925}, we have experimentally observed the same shift in a quantum interference pattern using entangled photon pairs enabling superior phase measurements that can, in principle, reach the Heisenberg limit.
Notably, the techniques we developed to control the coupling of Earth's rotation effect into our measurements are applicable to other photonics experiments  involving small and locally static natural quantities as gravitational fields. 
This experiment not only shows the potential of quantum photonics in precision measurements but opens up to the exploration of new frontiers in both quantum technology and fundamental physics research.
\\

\section*{Data availability}
All data pertaining to this study are available from the corresponding authors upon reasonable request.

\begin{acknowledgements}
The authors thank P.\,Schiansky, I.\,Agresti and  L.A.\,Rozema for help with the photon source and the photon detectors. 
The authors thank Z.\,Yin and H.\,Cao for discussions.
R.S. acknowledges support from Uni:Docs fellowship program, hosted by the University of Vienna.
H.Y. acknowledges support from the Marie-Sk{\l}odowska Curie Postdoctoral Fellowship program, hosted by the Horizon Europe.
C.H. and R.W.P. acknowledge support from the ESQ Discovery program (Erwin Schrödinger Center for Quantum Science and Technology), hosted by the Austrian Academy of Sciences (ÖAW).
P.W. acknowledges support in whole, or in part, from the research network TURIS, and by the European Union (ERC, GRAVITES, No 101071779). Views and opinions expressed are however those of the author(s) only and do not necessarily reflect those of the European Union or the European Research Council. Neither the European Union nor the granting authority can be held responsible for them. Further funding was received from the Austrian Science Fund (FWF) through BeyondC (F7113) and Research Group 5 (FG5). For the purpose of open access, the author has applied a CC BY public copyright license to any Author Accepted Manuscript version arising from this submission.
\end{acknowledgements}

\section*{Author contributions}
R.S., H.Y., and R.W.P. implemented the experiment and performed data analysis with leading input from R.S.
T.S. and C.H. provided help with the theoretical ideas and experimental implementation. T.S. assisted with the data analysis. R.W.P. conceived the experiment.
R.W.P., C.H., and P.W. supervised the project.
All authors contributed to the preparation of the manuscript, with leading input from H.Y.

\section*{Competing interests}
The authors declare no competing interests.

\bibliographystyle{apsrev4-2}
\bibliography{biblio}

\onecolumngrid
\PRLsep

\section*{Appendix A} 
\label{sec:Methods}
\noindent\textbf{Characteristics of the experimental setup:}
A periodically poled KTiOPO$_{4}$ crystal produces orthogonally-polarized photon pairs centered at \SI{1545.76}{\nano\meter} in a type-II spontaneous parametric down-conversion (SPDC) process. The crystal is pumped by continuous wave (CW) Ti:Sapphire laser (Coherent Mira HP) emitting at \SI{772.88}{\nano\meter} \cite{Greganti_2018}. 
The photon-source pump power is set to \SI{145}{\milli\watt}, leading to a detected photon-pair coincidence rate of approximately \SI{400}{\kilo\hertz}.
The photons in each generated pair are combined on a PBS, and are overlapped temporally using a delay line in one of the input ports of the PBS.
The total loss of the entire experimental setup is 90\% (\SI{10}{\dB}). The Sagnac loop introduces \SI{5}{\dB} of losses, out of which \SI{1}{\dB} is the optical switch insertion loss, \SI{1}{\dB} from the \SI{2}{\kilo\meter} polarization-maintaining (PM) fiber ($\approx\SI{0.5}{\decibel/\kilo\meter}$), and \SI{3}{\dB} from fiber connections, while the input and output of the optical setup contribute the remaining \SI{5}{\dB}.
In detection, photons from the output paths are coupled into single-mode fibers connected to superconducting nanowire single-photon detectors, housed in a \SI{1}{\kelvin} cryostat, with a detection efficiency of roughly \SI{95}{\percent} and a dark-count rate around \SI{300}{\hertz}.
Amplified detection signals are counted using a time tagging module with a timing resolution of \SI{156.25}{\pico\second}, and two-photon coincidence events are extracted using a coincidence window of \SI{3.75}{\nano\second}. In the two-photon N00N state measurements, when both photons are propagating through the interferometer, the detected photon pair rate is around \SI{4}{\kilo\hertz}, consistent with the expected exponential fragility to losses of a two-photon N00N state $1- \eta_{i}^{2}\approx99\%$, where $\eta_{\text{i}}=0.1$ is the total transmission efficiency of the interferometer.
In the one-photon measurements, one photon of the pair is used as a trigger while the other propagates through the interferometer. 
The total heralded single-photon rate in the two detection ports is around \SI{20}{\kilo\hertz}, which is compatible with the overall losses $1-\eta_{\text{t}}\eta_{\text{i}}\approx95\%$, where $\eta_{\text{t}}=0.5$ is the transmission coefficient of the trigger photon fiber path.
\\

\begin{figure*}[!hb]
\centering
\includegraphics[width=1\textwidth]{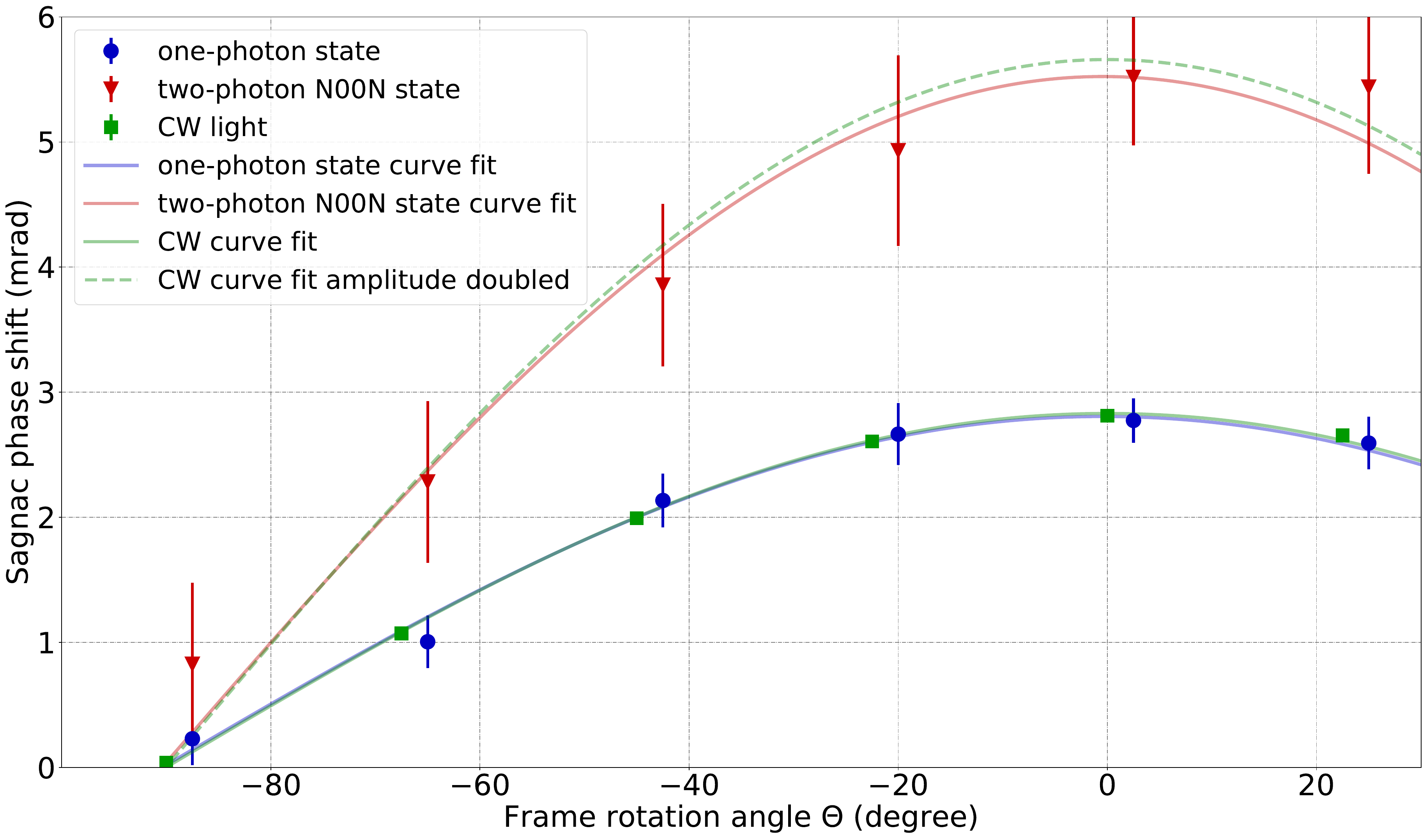}
\caption{\textbf{Comparison of photon measurements and CW measurements as calibration.}
Data of one-photon state (blue) and two-photon N00N state data (red) are inherited from Fig.~\ref{fig: frame rot} in Main text.
Green triangle marks are obtained using a polarimeter with CW light at six different $\Theta$'s ranging from -90° to +22.5°, evenly spaced by 22.5°. 
The green solid curve is the least-squares fit of CW measurements with fitting function as $\phi_{E}(\Theta) = S\Omega_{E}\cos(\Theta + \Theta_{0})$, 
This measurement allow us to find the frame angle that maximizes
the Sagnac phase ($\Theta$ = 0°).
The green dashed curve is plotted as $2S\Omega_{E}\cos(\Theta + \Theta_{0})$ to compare with the two-photon N00N state measurements.
}
\label{fig: cw_cal}
\end{figure*}

\noindent\textbf{Interferometer calibration:}
In the laboratory, the axis normal to fiber spool plane when vertically oriented with respect to the horizon is pointed north.
The rotational degree of freedom of the fiber loop frame introduces the opportunity for experimental calibration of the interferometer by estimating its scale factor S, while assuming Earth's rotation rate ($\Omega_{E}$) as a known quantity, with $\phi_s = S\Omega_{E}$. 
A set of phase measurements are performed with a CW light source at telecom wavelength at six different angular positions $\Theta^k$ of the fiber loop frame spaced by \SI{22.5}{\degree} (see Fig.~4), allowing us to find the frame angle that maximizes the Sagnac phase ($\Theta=0^\circ$).
H-polarized light is injected into the interferometer, which is converted into diagonal polarization by a half-wave plate before entering the Sagnac interferometer. Due to Earth's rotation, the H and V components acquire a relative Sagnac phase $\phi_{s}$, which is encoded in the polarization state ellipticity angle $\chi$~\cite{saleh2019fundamentals}, such that $\phi_{S} = 2\chi$.
A compact free-space polarimeter is employed to fully characterize the polarization state after the waveplates, which compensate first for the polarization rotation in the output fiber circulator path.
As in the measurements with quantum light, the optical switch is driven by a \SI{0.1}{\hertz} square-wave.
The recorded time trace of $\chi$ is partitioned into two sets by demodulating it using the driving signal. For each frame angle $\Theta^k$ the differential average between the two traces $\delta\overline{\chi}^k =\overline{\chi}_{on}^k-\overline{\chi}_{off}^k$ is used to calculate the Sagnac phase $\phi_S^k$ and its associated uncertainty $\sigma^k$.
As part of a Monte-Carlo simulation resampling the phase values $\phi_S^k$ using their uncertainties, the data are fit to the model function $\phi_{E}(\Theta) = S\Omega_{E}\cos(\Theta + \Theta_{0})$, where $\Omega_{E} = \SI{7.29e-5}{\radian\per\second}$ is the known value of Earth's rotation rate, and $S$ and $\Theta_{0}$ are free parameters. The Monte-Carlo simulation estimating these parameters additionally samples the frame angles $\Theta^k$ from uniform probability distributions $[\Theta^k-1,\Theta^k+1]$. 
The extracted fit parameters are the scale factor $S = \SI{38.8 \pm 0.1}{\second}$, and the angular offset $\Theta_{0}= \SI{0.03\pm0.33}{\deg}$. The CW measurements are compared with the photon measurements from Fig.~3 of the main text.
\\

\noindent\textbf{Noise mitigation.}
The interferometer frame is fixed on an air-floated optical table to dampen the transduction of ambient seismic vibrations into the frame. The fiber spools are covered with layers of insulation material (Thinsulate{\textregistered}) to mitigate temperature- and air-current-induced spatial gradients and time-varying fluctuations of the fiber length and refractive index. This passive isolation increases the scale factor stability in time by stabilizing the enclosed interferometric area. 
In addition, the optical switching method is also a fundamental and powerful tool in our experimental implementation. 
It not only provides a reference where the rotation effect, manifested as a non-reciprocal Sagnac phase, is absent, but also eliminates possible phase errors resulting from laser intensity fluctuations, imperfections in the input photon state, non-ideal polarization rotations during light propagation out of the fiber loop, and variations in mechanical stresses applied to the frame structure across its angular orientations.
Furthermore, the modulation of the signal at a specific frequency helps to mitigate slow frequency drifts in the measured phase via post-processing, thereby increasing its long-term stability over acquisition times spanning hours. 
\\

\noindent\textbf{Phase estimation and uncertainty analysis.}
The phase shifts and associated uncertainties presented in Fig.~\ref{fig: frame rot} of the main text are estimated using a Monte-Carlo simulation accounting for photon counting noise and uncertainties in the phase offset. In each round of the simulation the photon counts are sampled using a Poisson distribution with mean and standard deviation of $N_k^{s}$ and $\sqrt{N_k^{s}}$, respectively, where $N_k^{s}$ is the number of recorded photon counts for the offset phase $\phi_0^k$ and switch state $s\in\{\text{on},\text{off}\}$. Additionally, 
phase-offset noise, correlated between the on and off states, is sampled from a distribution derived from the waveplate motor repeatability and added to the offsets $\phi_0^k$. For each sampled data set a least-squares fit is performed, using the amplitude $N_0$, fringe visibility $\mathcal{V}$, and phase shift $\phi$ as free parameters. Finally, the values and uncertainties of these parameters are estimated using the mean and standard deviation, respectively, taken over $10^5$ repetitions of the simulation.


\begin{figure*}[!htb]
\centering
\includegraphics[width=1\textwidth]{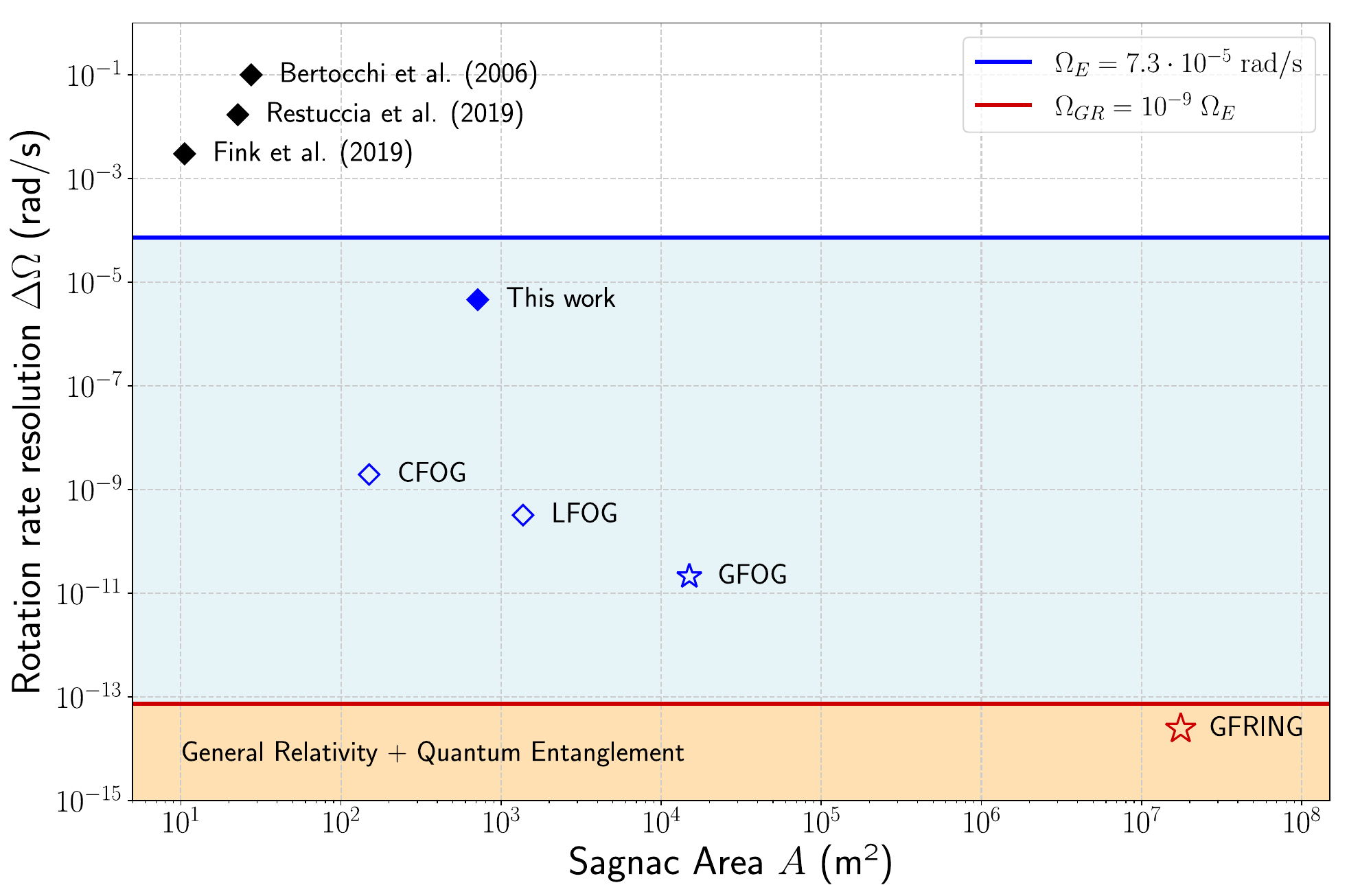}
\caption{\textbf{Rotation rate resolutions and enclosed areas of existing and predicted quantum optical Sagnac interferometers.}
The plot is divided into three sensitivity regimes: sensitivity below Earth rotation $\Omega_{E}$ (white zone), sensitivity above $\Omega_{E}$ but below general relativistic effects $\Omega_{GR}$ (blue zone), and sensitivity above $\Omega_{GR}$ (orange zone).
Diamond-shaped markers ($\blackdiam$, $\Diamond$) represent existing interferometric platforms, while star-shaped markers ($\medstar$) are proposed platforms but yet to be realized. Solid markers ($\blackdiam$) represent performed experiments with quantum states of light, while empty markers ($\Diamond$, $\medstar$) represent experiments yet to be performed. Bertocchi et al.~\cite{Bertocchi_2006}: $L_{f}=550~m, P=0.63~m$,  Restuccia et al.~\cite{Restuccia_2019}: $L_{f}=100~m, P=2.85~m$, Fink et al.~\cite{Fink_2019}: $L_{f}=270.5~m, P=0.49~m$, this work: $L_{f}=2~km, P=5.6~m$, CFOG (Commercial Fiber-Optic Gyroscope (FOG))~\cite{CFOG_2014}: $L_{f}=3~km, P=0.63~m$, LFOG (Large FOG)~\cite{LFOG_2020}: $L_{f}=8~km, P=2.15~m$, GFOG (Giant FOG)~\cite{GFOG_2017}: $L_{f}=15~km, P=12.57~m$, GFRING (Giant square Fiber RING interferometer): $L_{f}=47.5~km, P=6~km$, where $L_{f}$ is the fiber length and $P$ is the perimeter.
The photon-pair generation rate is \SI{1}{\giga\hertz} for the CFOG and LFOG, and \SI{10}{\giga\hertz} for the GFOG and GFRING, with integration times on the order of a month (for more details see Appendix B).}
\label{fig: exps_comparison}
\end{figure*}

\clearpage
	
\PRLsep

\section*{Appendix B}
\setcounter{equation}{0}
\renewcommand{\theequation}{S\arabic{equation}}
\section*{Interferometer with classical light}

\noindent \textbf{Continuous-wave (CW) light source:} The CW source employed is a broadband NIR-wavelength Superluminescent Diode (SLD). In front, a hard-coated bandpass filter with 12~nm full-width at half-maximum (FWHM) centered at 1545.5~nm is placed.
\\

\noindent\textbf{Interferometer operation:} The light enters horizontally polarized ($H$), is coupled into a single-mode (SM) fiber-optic circulator, and turns diagonally polarized ($+$) after a half-wave plate (HWP) oriented at 22.5° with respect to $H$. A Wollaston prism (WP) serves as a high-extinction ratio polarizing beam-splitter (PBS), which spatially separates the light into its horizontal and vertical ($V$) components. The output beams of the WP are coupled into polarization-maintaining (PM) fiber, aligned so that both polarization components propagate in the slow axis of the fiber. This ensures that the birefringence of the fiber does not contribute to the phase noise.
A four-port PM optical micro-electro-mechanical systems (MEMS) switch connects the two equal fiber spools of 1~km length. When this switch is in the state “ON”, the $H$ and $V$ components pick a relative non-reciprocal Sagnac phase shift $\phi_{S}$. 
After light recombination at the PBS, the polarization ellipse major axis turns to anti-diagonal (azimuth angle $\psi=-45^\circ$), while being slightly elliptical (ellipticity angle $\chi=\phi_{S}/2$). 
When passing the same HWP again, the polarization ellipse rotates by 45° with its major axis back to horizontal while still maintaining its ellipticity ($\psi=0, \chi=\phi_{S}/2$), then undergoes the fiber circulator and is coupled back to free space. It follows a set of 3 waveplates, respectively two quarter-wave plates (QWP) and one HWP, with a second WP as detection PBS (see Fig.~6)~\cite{Doerr_1994, Yao_2019}.
\\

\noindent\textbf{Waveplates operation:}
A set of two quarter-wave and one half-wave plate can be used to implement any polarization transformation~\cite{Simon_1990}. Thus, such a set of waveplates is able to perform several tasks at the same time.
\\
1) Polarization state tomography: They can select any polarization basis for the measurement, namely rectilinear $\{H,V\}$, diagonal $\{+,-\}$ ($\{D,A\}$) and circular $\{RL\}$, with their corresponding unitary operators to set the measurement basis being $\{I, \hat{\sigma}_{2}, \hat{\sigma}_{3}\}$. 
\\ 
2) Polarization compensation of the circulator fiber:
Since the SM fiber of the circulator introduces random birefringence, the polarization-state rotation at the output needs to be compensated for. This can be achieved by preparing the state before the fiber in two different known polarization states, for instance with two separate $H$ and $+$ polarizers. If full polarization tomography on the two output states $\ket{P_{H}}=U_{f}\ket{H}$, $\ket{P_{+}}=U_{f}\ket{+}$ is performed, the fiber unitary $U_{f}$ can be fully reconstructed. By configuring the waveplates to implement the inverse matrix $U^{-1}_{f}$ one reduces the overall action of the circulator on the polarization state to an identity transformation.
\\
3) Selecting the working point:
To set the working point of the interferometer it is sufficient to apply a unitary $U(\phi) = e^{-i\phi\hat{n}_{2}\cdot\vec{\sigma}/2}$ that rotates the compensated polarization state along the diagonal axis $\hat{n}_{2}$ by an angle $\phi$, which corresponds to the bias phase. The rotation operator can then simply be written as a phase shift matrix 
\begin{equation}
U(\phi) = \begin{pmatrix} 1 & 0\\0 & e^{i\phi} \end{pmatrix}
\end{equation}
expressed in the diagonal basis
\begin{equation}
U_\text{bias}(\phi)=U_\text{HWP}(-22.5^{\circ})U(\phi)U_\text{HWP}(-22.5^{\circ}),
\end{equation}
where $U_\text{HWP}(-22.5^{\circ})$ is the unitary of an HWP oriented at -22.5°. The operation physically corresponds to rotating the compensated $H$ polarization to $A$, applying a relative phase between the $H$ and $V$ components, and rotating back.
\\
In the end, the waveplates are able to implement an overall unitary $U = U_\text{proj}U_\text{bias}U^{-1}_{f}$, with $U_\text{proj} \in \{I, \hat{\sigma}_{2}, \hat{\sigma}_{3}\}$, by setting a triplet of angles $(\theta_{1},\theta_{2},\theta_{3})$ such that
\begin{equation}
U = U_\text{HWP}(\theta_{3})U_\text{QWP}(\theta_{2})U_\text{QWP}(\theta_{1}).
\end{equation}

\noindent\textbf{Phase extraction:} To measure the Sagnac phase encoded in the polarization state after the waveplates, a compact free-space polarimeter with a sampling rate of \SI{20}{\hertz} is employed. The phase is estimated as $\phi_{S}\approx2\sqrt{\delta\overline{\chi}^2+\delta\overline{\psi}^2}$, for small differential mean ellipticity $\delta\overline{\chi}$ and azimuth $\delta\overline{\psi}$ angles, where $\delta$ denotes the difference across the two switch states. Ideally, the Sagnac phase should only be encoded in the ellipticity angle. However, due to imperfect polarization rotation from the waveplates, the measurement bases will change, and a small part of the phase signal couples into the azimuth. The employed expression represents then a good approximation of the effective Sagnac phase value for small angular shifts.
\\

\noindent\textbf{Optical switch operation:}
A square-wave modulation at \SI{0.1}{\hertz} with a \SI{50}{\percent} duty cycle was employed for all the measurements (classical and quantum). Even though in principle our four-port PM fiber optical MEMS switch could operate at a modulation signal frequency of almost \SI{1}{\kilo\hertz}, we observed a noisy transition region of \SI{20}{\milli\second} duration centered at each rising/falling edges of the square signal. For this reason, in post-processing, photon counts within a \SI{10}{\milli\second} time window from to the edges are discarded to make sure none of the time tags fall in a noisy region, while for the CW measurement at least the two data samples around the transitions are removed (\SI{50}{\milli\second} time window) due to the limited sampling rate (\SI{20}{\hertz}) of the polarimeter (see Fig.~7). This procedure effectively limits the switching frequency to less than \SI{25}{\hertz}. In the end, a modulation at \SI{0.1}{\hertz} was selected since it showed the best stability over time. The optical switch also introduces a state-dependent power loss of \SI{10}{\percent} when is turned off, so that the transmission in the “OFF” state is 0.9 of the “ON” state.

\section*{Interferometer with quantum light}
\vspace{-6pt}
\noindent \textbf{Single-photon state measurement:}
A HWP oriented at 45° is placed into the $V$ photon input path, the photon is thus transmitted by the PBS and directly coupled into a single-mode fiber connected to a detector channel, serving as a trigger, while the H photon is transmitted out of the other port and injected into the interferometer (see Fig.~6). Two-photon coincidence counts between the two output channels and the trigger are recorded, and for each phase bias the total number of events of each switch state is $N^{(\text{on}/\text{off})}_{H}$.
The selected observable for the fringe scan points is the total normalized $V$ counts 
\begin{equation}
n_{V} = \frac{N_{V}}{N_{H}+N_{V}},
\end{equation}
where
\begin{align}
N_{H}&= \frac{A_{H}}{2}[1+ \mathcal{V}_{H}\cos(\phi + \phi^{(1)})],\\
N_{V}&= \frac{A_{V}}{2}[1- \mathcal{V}_{V}\cos(\phi - \phi^{(1)})]
\end{align}
with the constraints $\mathcal{V}_{H}=\mathcal{V}_{V}=\mathcal{V}$
and $\phi^{(1)}_{H}=\phi^{(1)}_{V}=\phi^{(1)}$. The corresponding fitting function is
\begin{equation}
n_{V}(\phi)=a_{V}\frac{1-\mathcal{V}\cos(\phi + \phi^{(1)})}{1+\eta \mathcal{V}\cos(\phi + \phi^{(1)})},
\end{equation}
where $a_{V}=A_{V}/(A_{H}+A_{V})$ and $\eta=(A_{H}-A_{V})/(A_{H}+A_{V})$. The fringe visibility $\mathcal{V}$, phase offset $\phi^{(1)}$ as well as the amplitudes $A_H$ and $A_V$ are taken as free parameters of the fit.
\\

\noindent \textbf{N00N state measurement:}
By removing or rotating the HWP to 0°, and adjusting the relative temporal delay to make the H and V photons indistinguishable in all other degrees of freedom, a $\ket{1_{H}}\ket{1_{V}}$ Fock state is generated after the input PBS.  The Hong-Ou-Mandel interference in polarization takes place at the 22.5° HWP, generating the two-photon N00N state. Coincidence photon counts between the two output channels are recorded, for each phase bias, and the total number of photons in the two switch states $N^{(\text{on}/\text{off})}_{HV}$ are counted. We select the latter quantity as observable for the fringe scan. The fitting function is
\begin{equation}
N_{HV}(\phi) = \frac{A_{HV}}{2}[1+\mathcal{V}\cos(2\phi + \phi^{(2)})],
\end{equation}
with $A_{HV}$, $\mathcal{V}$, and $\phi^{(2)}$ as free parameters.

\section*{Sensitivity analysis of proposed experiments:}
For the analysis of the proposed experiments with two-photon NOON states (empty markers), the Heisenberg phase precision scaling at the point of maximum sensitivity is assumed $\delta \phi=1/2\sqrt{R_{out}T}$, where $R_{out}$ is the detected photon pairs rate and T is the total integration time of the measurement. The newly proposed switching technique is included and $\delta \phi_{on}=\delta \phi_{off}=\delta \phi$ is imposed with $T/2$ integration time for each state, giving
\begin{equation}
\delta \phi=\frac{1}{\sqrt{2R_{out}T}}.
\end{equation}
The optical losses during propagation in fiber are exponential in its length, thus the output rate is related to the input rate as $R_{out}=\eta R_{in}$, with transmission coefficient $\eta = 10^{-\alpha L_{f}/10}$, where $\alpha$ and $L_{f}$ are the fiber attenuation coefficient and length, respectively. For a general NOON state the losses are exponential in the number of photons N such that $\eta\rightarrow\eta^{N}$, so the two-photon rate is
\begin{equation}
R_{out}=10^{-2\frac{\alpha L_{f}}{10}} R_{in}.
\end{equation}
In the end, the phase precision is given by 
\begin{equation}
\delta \phi=\frac{1}{\sqrt{2R_{in}T}} 10^{\frac{\alpha L_{f}}{10}}.
\end{equation}
The corresponding rotation rate precision is then given by the Sagnac phase formula at maximum signal $\delta \Omega=\delta \phi/S$, with the scale factor defined as $S = \frac{8\pi A}{\lambda c}$, where $\lambda$ is the light wavelength and $A$ is the effective area of the interferometer. $A=A(L_{f})$ is also a function of the fiber length depending on the interferometer geometry. In particular, for the proposed giant square ring interferometer (GFRING) the effective area is $A=\frac{1}{n_t}(\frac{L_{f}}{4})^{2}$, where $n_{t}$ are the number of fiber turns around the square frame. Given the expected dimensions of the ring it is assumed that its frame plane would be fixed parallel to the Earth surface at its location. The rotation rate resolution can then be written as
\begin{equation}
\delta \Omega=\sqrt{\frac{2}{{R_{in}T}}} \frac{\lambda c}{\pi \sin{\theta_{L}}} n_{t} \frac{10^{\frac{\alpha L_{f}}{10}}}{{L_{f}^2}},
\end{equation}
where $\theta_{L}$ is the latitude at the location of the ring.
The geometrical parameters of the GFRING are numerically extracted by requiring the signal-to-noise ratio (SNR) to be $\Omega_{GR}/\delta \Omega = 3$ and maximizing $n_{t}$ to reduce the surface area of the ring as much as possible. The calculation assumes the latitude angle $\theta_{L}=48.2^{\circ}$ of Vienna, Austria. Finally, the optimal fiber length and number of turns result in $L_{f}=47.5$~km and $n_{t}=8$, corresponding to a square side of 1.5~km and a perimeter of 6~km. For each experiment an integration time of $T=2$ months is considered (see Table 3).


\begin{figure*}[!thb]
\centering
\includegraphics[width=\textwidth]{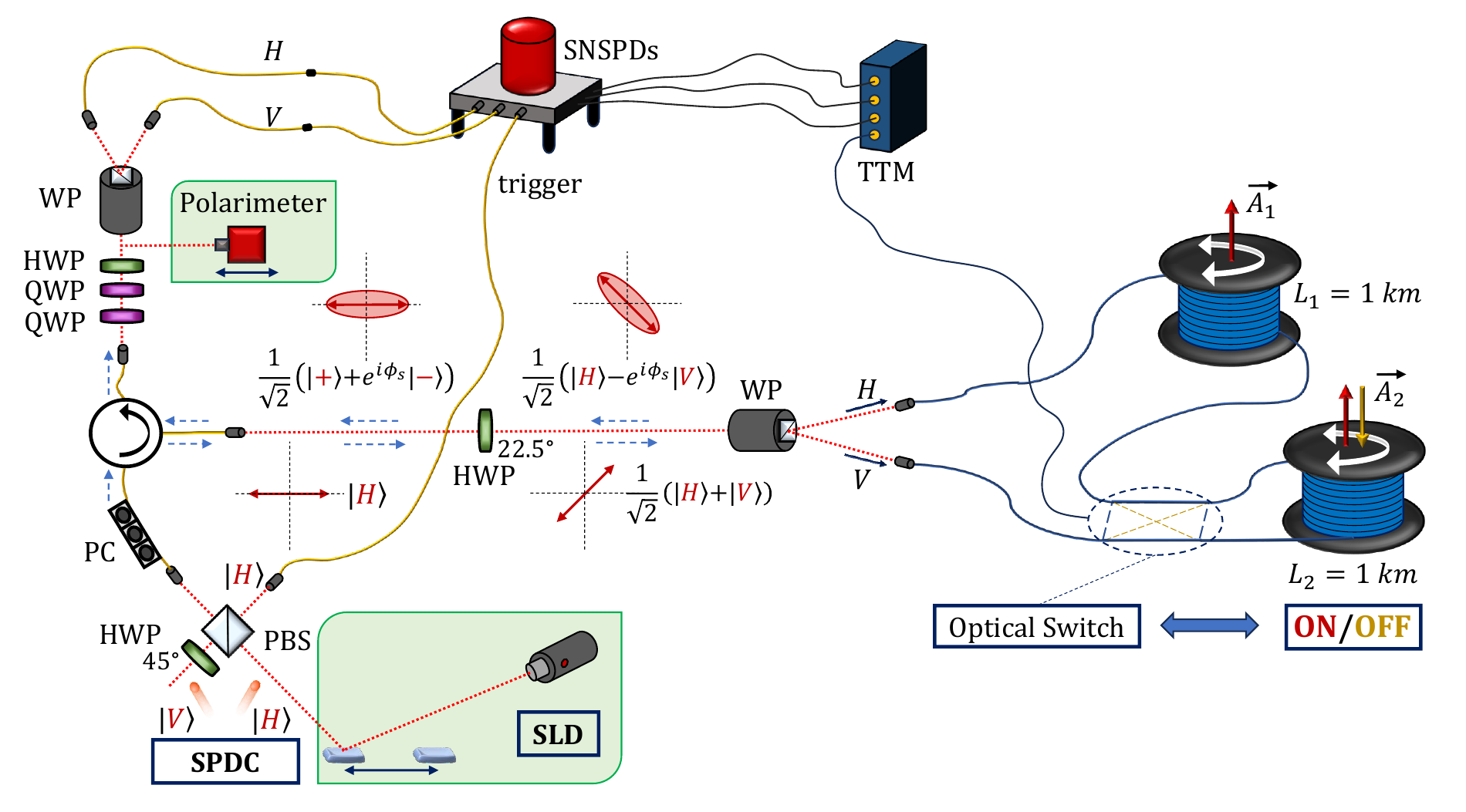}
\caption{\textbf{Detailed experimental setup.} Experimental scheme for the single-photon and CW measurements (green insets). In the one-photon measurement a HWP at \SI{45}{\degree} before the input PBS is used to send the trigger photon directly to the SNSPDs. For the CW measurement the SNSPDs are replaced with a polarimeter placed after the waveplates. The polarization states before/after the Sagnac are indicated with red arrows, and the Earth rotation phase manifests as ellipticity in the output polarization state. \textbf{SPDC.} Spontaneous Parametric Down-Conversion single photons source. \textbf{SLD.} SuperLuminescent Diode. \textbf{PBS.} Polarizing Beam Splitter cube. \textbf{PC.} fiber Polarization Controller. \textbf{WP.} Wollaston Prism. \textbf{HWP.} Half-wave plate. \textbf{QWP.} Quarter-wave plate. \textbf{SNSPDs.} Superconducting Nanowire Single-Photon Detectors. \textbf{TTM.} Time Tagging Module.
}  
\label{fig: exp_setup}
\end{figure*}

\begin{figure*}[!thb]
\centering
\includegraphics[width=\textwidth]{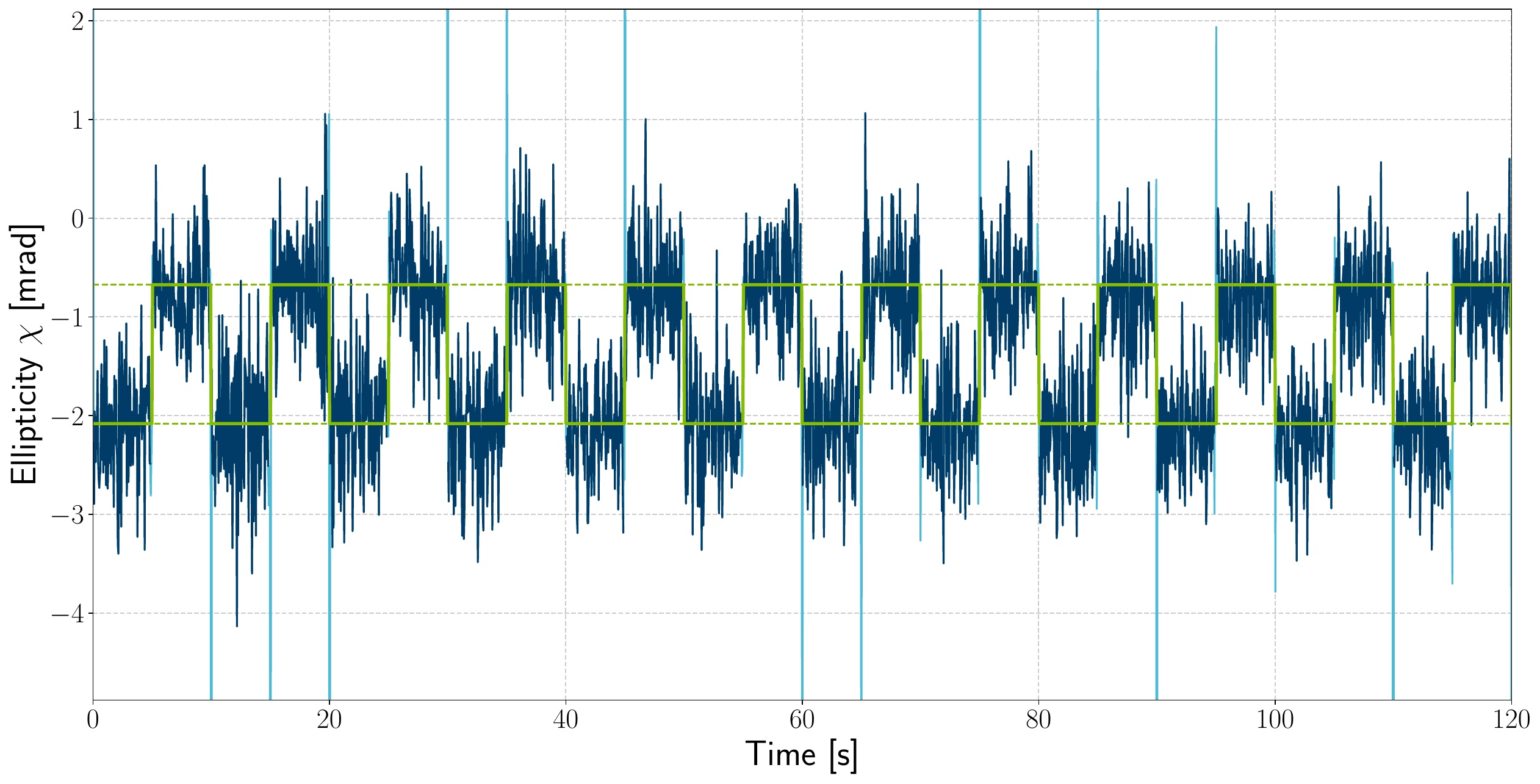}
\caption{\textbf{Ellipticity angle time signal.} The ellipticity angle measured with CW light by a polarimeter, with \SI{20}{\hertz} sampling rate, is plotted over time. It is shown how the square-wave voltage applied to the switch at \SI{0.1}{\hertz} modulates the measured signal (dark blue trace). The high/low levels correspond to the switch on/off states (green square wave). Data samples in the trace are cut around the transition points (light blue samples) and two separate on/off time traces are extracted. $\delta\overline{\chi}$ is calculated as the difference between the two respective averages $\overline{\chi}_{on} - \overline{\chi}_{off}$ (green dashed horizontal lines). The same analysis is performed on the azimuth time trace.}
\label{fig: ellipticity switch}
\end{figure*}

\FloatBarrier
\section*{Tables} 
\begin{table}[thb]
\begin{minipage}{\textwidth}
\caption{\textbf{Single-photon state measurement extracted parameters.}}\label{tab1}%
\begin{tabular}{ p{0.1\linewidth}|p{0.15\linewidth}p{0.15\linewidth}p{0.15\linewidth}p{0.15\linewidth}p{0.15\linewidth} }
\toprule
$\Theta$ & $\mathcal{V}_\text{on}$ & $\mathcal{V}_\text{off}$ & $\phi_\text{on}$ (mrad) & $\phi_\text{off}$ (mrad) & $\phi_\text{E}$ (mrad)\\
\midrule
$-87.5^\circ$   & 99.66(12) \%  & 99.67(12) \%   & -1.36(2.45)   & -1.13(2.45)    & 0.23(21)\\
$-65^\circ$     & 99.68(12) \%  & 99.69(12) \%   & -1.99(2.45)   & -0.98(2.45)    & 1.00(21)\\
$-42.5^\circ$   & 99.69(12) \%  & 99.68(12) \%   & -1.86(2.45)   & ~\,0.28(2.45)  & 2.14(21)\\
$-20^\circ$     & 99.59(13) \%  & 99.60(13) \%   & -7.80(2.45)   & -5.13(2.45)    & 2.66(25)\\ 
$~~\,2.5^\circ$ & 99.69(13) \%  & 99.67(13) \%   & -2.43(2.43)   & ~\,0.34(2.46)  & 2.77(18)\\
$~~\,25^\circ$  & 99.66(12) \%  & 99.66(12) \%   & -1.22(2.45)   & ~\,1.37(2.45)  & 2.59(21)\\
\bottomrule
\end{tabular}
\footnotetext{Interferometric visibility $\mathcal{V}_{\text{on}/\text{off}}$ and phase offset $\phi_{\text{on}/\text{off}}$ values extracted from the fit for each frame orientation angle $\Theta$ both in the switch on and off states, with the corresponding Earth rotation induced phases $\phi_\text{E}$ calculated as $\phi_\text{off}-\phi_\text{on}$.}
\end{minipage}
\end{table}



\begin{table}[thb]
\begin{minipage}{\textwidth}
\caption{\textbf{Two-photon NOON state measurement extracted parameters.}}\label{tab2}
\begin{tabular}{ p{0.1\linewidth}|p{0.15\linewidth}p{0.15\linewidth}p{0.15\linewidth}p{0.15\linewidth}p{0.15\linewidth} }
\toprule
$\Theta$ & $\mathcal{V}_\text{on}$ & $\mathcal{V}_\text{off}$ & $\phi_\text{on}$ (mrad) & $\phi_\text{off}$ (mrad) & $\phi_\text{E}$ (mrad)\\
\midrule
$-87.5^\circ$   & 96.81(45) \%  & 96.79(45) \%   & -28.53(4.94)      & -27.71(4.95)      & 0.82(65)\\
$-65^\circ$     & 96.78(45) \%  & 96.73(45) \%   & -21.76(4.93)      & -19.49(4.94)      & 2.28(65)\\
$-42.5^\circ$   & 96.72(45) \%  & 96.56(45) \%   & ~\,\,-9.87(4.94)  & ~\,\,-6.02(4.95)  & 3.86(65)\\
$-20^\circ$     & 95.67(44) \%  & 95.56(44) \%   & ~\,\,-5.00(4.95)  & ~\,\,-0.07(4.96)  & 4.93(76)\\
$~~\,2.5^\circ$ & 97.14(45) \%  & 97.17(45) \%   & -24.60(4.92)      & -19.09(4.92)      & 5.51(54)\\
$~~\,25^\circ$  & 97.54(46) \%  & 97.51(46) \%   & -10.53(4.94)      & ~\,\,-5.09(4.95)  & 5.44(69)\\
\bottomrule
\end{tabular}
\end{minipage}
\end{table}

\begin{table}[thb]
\begin{minipage}{\textwidth}
\caption{\textbf{Quantum optical Sagnac interferometers specifications and resolutions comparison.}}\label{tab3}%
\begin{tabular}{ c@{\hspace{3pt}} l|p{0.1\linewidth}p{0.1\linewidth}p{0.1\linewidth}p{0.1\linewidth}p{0.15\linewidth}p{0.1\linewidth} }
\toprule
& & $L_{f}$ & $P$ & $A$ & $S~(\text{s}^{-1})$ & $\delta\phi_{S}$ (rad) & $\delta\Omega~(\text{rad s}^{-1})$\\
\midrule
$\squaremarker$ &Our work & 2~km    & 5.55~m   & $715~\text{m}^{2}$   & 38.8   & $1.79\cdot10^{-4}$ & $4.61\cdot10^{-6}$\\
$\circmarker$   &CFOG     & 3~km    & 0.63~m   & $150~\text{m}^{2}$   & 8.1    & $1.34\cdot10^{-8}$ & $1.95\cdot10^{-9}$\\
$\circmarker$   &LFOG     & 8~km    & 2.15~m   & $1372~\text{m}^{2}$  & 74     & $2.38\cdot10^{-8}$ & $3.21\cdot10^{-10}$\\
$\circmarker$   &GFOG     & 15~km   & 12.57~m  & $15000~\text{m}^{2}$ & 811    & $1.69\cdot10^{-8}$ & $2.11\cdot10^{-11}$\\ 
$\squaremarker$ &GFRING   & 47.5~km & 6~km     & $17.6~\text{km}^{2}$ & 951320 & $2.31\cdot10^{-8}$ & $2.43\cdot10^{-14}$\\
\bottomrule
\end{tabular}
\footnotetext{Square markers ($\squaremarker$) represent square interferometer frames, while ($\circmarker$) are circular frames. The selected optical wavelength $\lambda$ is \SI{1550}{\nano\meter} for each proposed experiment. The existing and predicted FOGs platforms utilize PM fibers ($\alpha=0.5$~dB/km), while for the GFRING a standard single-mode fiber ($\alpha=0.16$~dB/km) is proposed. The single-photon pair generation rate $R_{in}$ is 1~GHz, while for the GFOG and GFRING is 10~GHz. $L_{f}$ is the fiber length, $P$ is the frame perimeter, $A$ is the effective interferometer area, $S$ is the scale factor, while $\delta \phi_{S}$ and $\delta \Omega$ are the Sagnac phase and rotation resolutions.}
\end{minipage}
\end{table}
\end{document}